\documentclass[twocolumn,secnumarabic,amssymb, amsmath, nofootinbib,tightenlines,
nobibnotes, aps, prl,epsfig]{revtex4}
\usepackage{graphicx}
\usepackage{dcolumn}
\usepackage{bm}
\begin{document}
\preprint{APS/123-QED}
\title{Analytic derivation of the non-linear gluon distribution function }

\author{G.R.Boroun}%
 \email{grboroun@gmail.com; boroun@razi.ac.ir }
\affiliation{Department of Physics, Razi University, Kermanshah
67149, Iran}
\date{\today}
\begin{abstract}
In the present article, two analytical solutions based on the
Laplace transforms method for the linear and non-linear gluon
distribution functions have been presented at low values of $x$.
These linear and non-linear methods are presented based on the
solutions of the Dokshitzer-Gribov- Lipatov-Altarelli-Parisi
(DGLAP) evolution equation and the Gribov-Levin-Ryskin Mueller-Qiu
(GLR-MQ) equation at the leading-order accuracy in perturbative
QCD respectively. The gluon distributions are obtained directly in
terms of  the parametrization of structure function
$F_{2}(x,Q^{2})$ and its derivative and compared with the results
from the parametrization models. The $n_{f}$ changes at the
threshold are considered in the numerical results. The effects of
the non-linear corrections are visible as $Q^{2}$ decreases and
vanish as $Q^{2}$ increases. The nonlinear corrections tame the
behavior of the gluon distribution function at low
 $x$ and $Q^{2}$ in comparison with the parametrization models.\\
\end{abstract}
 \pacs{***}
\maketitle
\subsection{I.Introduction}

In recent years,  the study and consider of solutions of the
Dokshitzer-Gribov- Lipatov-Altarelli-Parisi (DGLAP) evolution
equations [1] based on Laplace transforms method have been
considered by many authors [2-4]. Firstly authors in Ref.[5]
showed that it is possible to solve the leading order (LO) DGLAP
evolution equations directly from the parametrization of the
proton structure function $F_{ 2}(x,Q^{2})$ based on Laplace
transforms method. These evolution equations in Ref.[5] are a set
of integro-differential equations which can be used to evolve the
quark and gluon distributions. The parton distribution functions
in hadrons play a key role in the Standard Model processes and
searches for new physics in future accelerators. The structure
function of the proton measured experimentally in deep inelastic
scattering processes then traditionally the gluon and quark
distribution functions where have been determined simultaneously
by fitting experimental data  on the proton structure function at
small values of the Bjorken variable $x$. At LO approximation the
proton structure function is expressed through the quark density
as
$F_{2}(x,Q^{2})=\sum_{i=1}^{n_{f}}{e_{i}^{2}}x[q(x,Q^{2})+\overline{q}(x,Q^{2})]$,
where $n_{f}$ is the number of flavors.\\
Authors in Ref.[5] derived an explicit expression for the gluon
distribution function $G(x,Q^{2})=xg(x,Q^{2})$ in the proton in
terms of the proton structure function $F_{ 2}(x,Q^{2})$ by
solving the LO DGLAP equation for the $Q^{2}$ evolution of $F_{
2}(x,Q^{2})$  analytically. In particular, accurate knowledge of
gluon distribution functions at small $x$ and small $Q^{2}$ will
play a vital role in the electron-proton future colliders. Indeed,
the non-linear corrections (NLC) play an important role in the
small $x$ and small $Q^{2}$ regions at the Large Hadron electron
Collider (LHeC) and Future Circular Collider hadron-electron
(FCC-he) [6]. The non-linear corrections of the gluon
recombination to the parton distributions have been calculated by
Gribov-Levin- Ryskin (GLR) and Mueller-Qiu (MQ) in [7] based on
the Abramovsky-Gribov-Kancheli (AGK) cutting rules in the double
leading logarithmic approximation (DLLA). It is known that the
gluon recombination effects reduce the growth of the gluon
distribution function, therefore cannot be negligible at the small
$x$ and $Q^{2}$ regions. Indeed all possible $g+g{\rightarrow}g$
ladder recombinations are resummed to leading order of the
parameter $\alpha_{s}\ln(1/x)\ln(Q^{2}/Q^{2}_{0})$ where leads to
saturation
of the gluon density at this region.\\
In the following sections, we present two analytic methods where
determine $G(x,Q^{2})$ directly from $F_{2}^{p}(x,Q^{2})$ and its
derivative into $\ln{Q^{2}}$. The first one is a review for the
linear evolution equations based on the Laplace transforms method
for the DGLAP evolution equation. The second one is the same
method for the GLR-MQ equation which presents the non-linear
corrections to the gluon distribution function directly from the
proton structure function and its derivative. These methods lead
to equivalent results without the intervening
differential equation.\\
The non-linear corrections emerge from the recombination of two
gluon ladders where modify the evolution equation of singlet quark
distribution by an extra  non-linear term. This non-linear term
add to the linear DGLAP evolution equation by the following form
\begin{eqnarray}
\frac{\partial{F_{2}(x,Q^{2})}}{\partial{\ln}Q^{2}}&=&\frac{\partial{F_{2}(x,Q^{2})}}{\partial{\ln}Q^{2}}|_{DGLAP}\\
&&-<e^{2}>\frac{27\alpha_{s}^{2}(Q^{2})}{160\mathcal{R}^{2}Q^{2}}[xg(x,Q^{2})]^{2}+HT,\nonumber
\end{eqnarray}
where
\begin{eqnarray}
\frac{\partial{F_{2}(x,Q^{2})}}{\partial{\ln}Q^{2}}|_{DGLAP}&=&\frac{\alpha_{s}(Q^{2})}{4\pi}[P_{qq}(x){\otimes}F_{2}(x,Q^{2})\nonumber\\
&&+<e^{2}>P_{qg}(x){\otimes}G(x,Q^{2})]
\end{eqnarray}
The splitting functions $P_{ij}$ are the Altarelli-Parisi
splitting kernels at one loop correction, and $<e^{2}>$ is the
average of the charge $e^{2}$ for the active quark flavors,
$<e^{2}>=n_{f}^{-1}\sum_{i=1}^{n_{f}}e_{i}^{2}$. We take the
$n_{f}=3$ for $\mu^{2}<m^{2}_{c}$, $n_{f}=4$ for
$m^{2}_{c}<\mu^{2}<m^{2}_{b}$ and $n_{f}=5$ for
$\mu^{2}>m^{2}_{b}$ and adjust the QCD parameter $\Lambda$ at each
heavy quark mass threshold. The correlation length $\mathcal{R}$
determine the size of the non-linear term. This value depends on
how the gluon ladders are coupled to the nucleon or on how the
gluons are distributed within the nucleon. The $\mathcal{R}$ is
approximately equal to $\simeq 5~\mathrm{GeV}^{-1}$ if  gluons are
populated across the proton and it is equal to $\simeq
2~\mathrm{GeV}^{-1}$ if gluons have the hotspot like structure.
Here the higher dimensional gluon
distribution(i.e., higher twist) is assumed to be zero.\\
Now we review the method of extracting the gluon distribution from
the parametrization of the proton structure function and its
derivative in the  linear and non-linear corrections using the
Laplace transforms method in next sections respectively.\\

\subsection{II. Linear Formalism}

By considering the variable changes $\nu{\equiv}\ln(1/x)$ and
$w{\equiv}\ln(1/y)$, one can rewrite the DGLAP evolution equation
(i.e., Eq.2) in $s$-space as
\begin{eqnarray}
\frac{\partial{f_{2}(s,Q^{2})}}{\partial{\ln}Q^{2}}&=&
\Phi_{f}(s,Q^{2})f_{2}(s,Q^{2})\nonumber\\
&&+<e^{2}>\Theta_{f}(s,Q^{2})g(s,Q^{2}),
\end{eqnarray}
where the Laplace-transform of the distribution functions read
\begin{eqnarray}
{\mathcal{L}}[\mathcal{\widehat{F}}_{2}(\upsilon,Q^{2});s]&=&f_{2}(s,Q^{2}),\nonumber\\
{\mathcal{L}}[{\widehat{G}}(\upsilon,Q^{2});s]&=&g(s,Q^{2})
\end{eqnarray}
where
\begin{eqnarray}
\mathcal{\widehat{F}}_{2}(\upsilon,Q^{2})&=&
F_{2}(e^{-\nu},Q^{2}),\nonumber\\
{\widehat{G}}(\upsilon,Q^{2})&=&G(e^{-\nu},Q^{2}).
\end{eqnarray}
and the coefficient functions $\Phi$ and $\Theta$ in $s$-space are
given by
\begin{eqnarray}
\Theta_{f}(s,Q^{2})&=&n_{f}\frac{\alpha_{s}(Q^{2})}{2\pi}(\frac{1}{1+s}-\frac{2}{2+s}+\frac{2}{3+s}),\nonumber\\
\Phi_{f}(s,Q^{2})&=&\frac{\alpha_{s}(Q^{2})}{4\pi}[4-\frac{8}{3}(\frac{1}{1+s}+\frac{1}{2+s}\nonumber\\
&&+2S(s))].
\end{eqnarray}
In the above equation the quantity $S(s)$ is related with the
Euler $\Psi(s+1)$ function as $S(s)=\Psi(s+1)+\gamma_{E}$ where
$\Psi(s)$ is defined by $\Psi(s)=\frac{d}{ds}{\ln}\Gamma(s)$. Here
$\Psi(x)$ is the digamma function and $\gamma_{E}=0.5772156 . .$
is Euler constant. We introduce the notion of the so-called nested
sums [8] throughout the rest of the paper where the function
$S(s)$ is defined
\begin{eqnarray}
S(s)=-\ln(2)-\sum_{l=0}^{\infty}\frac{(-1)^{l+1}}{s+l+1}.
\end{eqnarray}
\begin{figure}[h]
\includegraphics[width=0.45\textwidth]{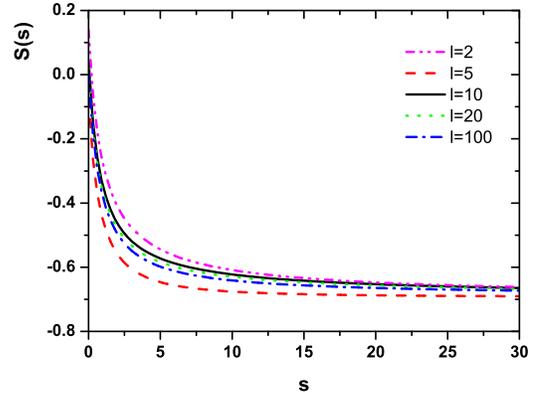}
\caption{Sensitivity of the function $S(s)$ verses $s$ for
different values of $l$.}\label{Fig1}
\end{figure}
In Fig.1 we consider the expansion of Eq.(7) by using different
points. We observe that for $l\geq 10$, the results are almost
equivalent and have the smooth behavior  for $s>10$. In this
article we shall widely use the notation $l=10$.\\
The LO solution of the gluon distribution in $s$-space in Eq.(3)
reads
\begin{eqnarray}
g(s,Q^{2})&=&k(s,Q^{2})Df_{2}(s,Q^{2})-h(s,Q^{2})f_{2}(s,Q^{2}),
\end{eqnarray}
where
\begin{eqnarray}
Df_{2}(s,Q^{2})&=&{\partial{f_{2}(s,Q^{2})}}/{\partial{\ln}Q^{2}},\nonumber\\
k(s,Q^{2})&=&1/(<e^{2}>\Theta_{f}(s,Q^{2})),\nonumber\\
h(s,Q^{2})&=&\Phi_{f}(s,Q^{2})/(<e^{2}>\Theta_{f}(s,Q^{2})).
\end{eqnarray}
The inverse Laplace transform of coefficients $k$ and $h$ in
Eq.(9) are defined by the following forms
\begin{widetext}
\begin{eqnarray}
k(\nu,Q^{2})&{\equiv}&{\mathcal{L}}^{-1}[k(s,Q^{2});\nu]=\frac{\pi}{2<e^2>\alpha_{s}}\{\delta'(\nu)+3\delta(\nu)-\exp(-\frac{3}{2}\nu)
[2\cos(\frac{1}{2}\sqrt{7}\nu)+\frac{6}{7}\sqrt{7}\sin(\frac{1}{2}\sqrt{7}\nu)]\},\nonumber\\
h(\nu,Q^{2})&{\equiv}&{\mathcal{L}}^{-1}[h(s,Q^{2});\nu]=\frac{1}{<e^2>}\{-(\frac{1}{2}+\frac{2}{3}\ln{2})\delta'(\nu)-(\frac{1}{6}+2\ln{2})\delta(\nu)
+\exp(-\frac{3}{2}\nu)[3.606\cos(\frac{1}{2}\sqrt{7}\nu)\nonumber\\
&&+1.371\sin(\frac{1}{2}\sqrt{7}\nu)]+\frac{1}{2}\exp(-4\nu)-\frac{8}{7}\exp(-5\nu)
+\frac{20}{11}\exp(-6\nu)-\frac{5}{2}\exp(-7\nu)+\frac{35}{11}\exp(-8\nu)\nonumber\\
&&-\frac{112}{29}\exp(-9\nu)+\frac{168}{37}\exp(-10\nu)-\frac{120}{23}\exp(-11\nu)\}.
\end{eqnarray}
\end{widetext}
Finally the gluon distribution function directly is obtained from
the parameterization of the structure function $F_{2}(x,Q^{2})$
and its derivatives by the following form
\begin{widetext}
\begin{eqnarray}
G(x,Q^{2})&=&\frac{\pi}{2<e^2>\alpha_{s}}\{\frac{\partial{DF_{2}(x,Q^{2})}}{\partial{\ln{x}}}+3DF_{2}(x,Q^{2})-\int_{x}^{1}\frac{dy}{y}DF_{2}(y,Q^{2})(\frac{x}{y})^{\frac{3}{2}}
[2\cos(\frac{1}{2}\sqrt{7}\ln(\frac{y}{x}))+\frac{6}{7}\sqrt{7}\sin(\frac{1}{2}\sqrt{7}\ln(\frac{y}{x}))]\}\nonumber\\
&&-\frac{1}{<e^2>}\{(\frac{1}{2}+\frac{2}{3}\ln{2})\frac{\partial{F_{2}(x,Q^{2})}}{\partial{\ln{x}}}+(\frac{1}{6}+2\ln{2})F_{2}(x,Q^{2})-\int_{x}^{1}\frac{dy}{y}F_{2}(y,Q^{2})(\frac{x}{y})^{\frac{3}{2}}
[3.606\cos(\frac{1}{2}\sqrt{7}\ln(\frac{y}{x}))\nonumber\\
&&+1.371\sin(\frac{1}{2}\sqrt{7}\ln(\frac{y}{x}))]
-\int_{x}^{1}\frac{dy}{y}F_{2}(y,Q^{2})(\frac{1}{2}(\frac{x}{y})^{4}
-\frac{8}{7}(\frac{x}{y})^{5}+.....-\frac{120}{23}(\frac{x}{y})^{11})\}.
\end{eqnarray}
\end{widetext}
The parameterization of the structure function $F_{2}(x,Q^{2})$,
which describes fairly well the available experimental data [9] on
the reduced cross sections in a full accordance with the Froissart
predictions in a range of the kinematical variables $x$ and
$Q^{2}$, $x{\leq}0.1$ and
$0.15~\mathrm{GeV}^{2}<Q^{2}<3000~\mathrm{GeV}^{2}$, suggested by
authors in Ref.[10]. This parametrization and its derivative read
\begin{eqnarray}
F_{ 2}(x,Q^{2})& =& D(Q^{2})(1-
x)^{n}\sum_{m=0}^{2}A_{m}(Q^{2})L^{m},
\end{eqnarray}
and
\begin{eqnarray}
DF_{2}(x,Q^{2}){\equiv}\frac{{\partial}F_{2}(x,Q^{2})}{\partial{\ln}Q^{2}}&=&
F_{2}(x,Q^{2})[\frac{{\partial}{\ln}D(Q^{2})}{\partial{\ln}Q^{2}}\nonumber\\
&&+\frac{{\partial}{\ln}\sum_{m=0}^{2}A_{m}(Q^{2})L^{m}}{\partial{\ln}Q^{2}}],\nonumber
\end{eqnarray}
where the effective parameters are defined in Appendix A and Table
I. Consequently, one can obtain the gluon distribution into the
effective coefficients obtained from a combined fit of the H1 and
ZEUS collaborations data.\\

\subsection{III. Non-Linear Formalism}

Now, the above procedure is used to derive the non-linear
corrections to the  gluon distribution function directly from the
parametrization of the proton structure function and its
derivatives. The resulting modified structure function should now
be driven by non-linear DGLAP evolution in $s$-space in a limited
approach. In $\nu$-space, we have defined the Laplace transform of
the ${\mathcal{L}}[\widehat{G}^{2}(\nu,Q^{2});s]$ to be less than
$g^{2}(s,Q^{2}){\equiv}[G(s,Q^{2})]^2$. Indeed
${\mathcal{L}}[\widehat{G}^{2}(\nu,Q^{2});s]<{\mathcal{L}}[\widehat{G}(\nu,Q^{2});s]^{2}
$.\footnote{The standard parametrization  of the gluon
distribution function at low $x$ introduced by $$
G(x,Q^{2})=f(Q^{2})x^{-\delta}
$$ where the low $x$ behavior could well be more singular. By considering the variable change
$\nu{\equiv}\ln(1/x)$, one can rewrite the gluon distribution in
$s$-space as
\begin{eqnarray}
{\mathcal{L}}[\widehat{G}^{2}(\nu,Q^{2});s]{=}\frac{f(Q^{2})^{2}}{(s-2\delta)},\nonumber\\
{\mathcal{L}}[\widehat{G}(\nu,Q^{2});s]^{2}{=}\frac{f(Q^{2})^{2}}{(s-\delta)^{2}}.\nonumber
\end{eqnarray}
We observe that the function
${\mathcal{L}}[\widehat{G}^{2}(\nu,Q^{2});s]$  is always lower
than ${\mathcal{L}}[\widehat{G}(\nu,Q^{2});s]^{2}$ for low $s$
values in a wide range of $Q^{2}$ values. According to this
result, we use from this limited approach for solving the
quadratic equation in $s$-space.}
Therefore in this limit, we take
the Laplace transform of (1), by follow
\begin{eqnarray}
\frac{\partial{f_{2}(s,Q^{2})}}{\partial{\ln}Q^{2}}&{\simeq}&
\Phi_{f}(s)f_{2}(s,Q^{2})+<e^{2}>\Theta_{f}(s)g(s,Q^{2})\nonumber\\
&&-<e^{2}>\zeta g^{2}(s,Q^{2}),
 \end{eqnarray}
where
$\zeta=\frac{27\alpha_{s}^{2}(Q^{2})}{160\mathcal{R}^{2}Q^{2}}$.
The non-linear gluon distribution function is defined by a
quadratic equation in $s$-space in the following form
\begin{eqnarray}
g^{2}(s,Q^{2})-k(s,Q^{2})g(s,Q^{2})+h(s,Q^{2})=0,
\end{eqnarray}
where
\begin{eqnarray}
h(s,Q^2)&=&\frac{1}{<e^{2}>\zeta}[Df_{2}(s,Q^{2})-\Phi_{f}(s,Q^{2})f_{2}(s,Q^{2})],\nonumber\\
k(s,Q^{2})&=& \Theta_{f}(s,Q^{2})/\zeta.
\end{eqnarray}
One can easily solve this equation and extract the non-linear
gluon distribution in $s$-space as
\begin{eqnarray}
g(s,Q^{2})=\frac{1}{2}k(s,Q^{2})[1{\pm}(1-\frac{4h(s,Q^{2})}{k^{2}(s,Q^{2})})^{1/2}
].
\end{eqnarray}
The quadratic equation (16) has two roots, which the negative root
reads
\begin{eqnarray}
g(s,Q^{2})&=&\frac{h(s,Q^{2})}{k(s,Q^{2})}[\mathrm{Linear~
Term}]+\{\frac{h^{2}(s,Q^{2})}{k^{3}(s,Q^{2})}\nonumber\\
&&+2\frac{h^{3}(s,Q^{2})}{k^{5}(s,Q^{2})}+...\}[\mathrm{Non-Linear~
Terms}].\nonumber\\
\end{eqnarray}
The non-linear terms in Eq.(17) are coefficients in the form
$\sum_{n=1}\zeta^{n}$, where $\zeta$ is around the order
$\mathcal{O}(\sim10^{-3})$ at $Q=1~\mathrm{GeV}$ and
$\mathcal{R}=2~\mathrm{GeV}^{-1}$, so this series is convergent
when $n{\rightarrow}\infty$.\\
We now re-derive our analytic solution using the inverse Laplace
transforms method for $g(s,Q^{2})$(i.e., Eq.(17)). The inverse
Laplace transform of terms in Eq.(17) are given by the following
form
\begin{eqnarray}
G^{\mathrm{NLC}}(x,Q^{2})&=&\mathrm{Eq}.(10)+
{\mathcal{L}}^{-1}[\frac{h^{2}(s,Q^{2})}{k^{3}(s,Q^{2})}\nonumber\\
&&+2\frac{h^{3}(s,Q^{2})}{k^{5}(s,Q^{2})}+...;\nu].
\end{eqnarray}
Therefore the non-linear corrections to the gluon distribution
function due to the Laplace transforms method is defined directly
from the parametrization of the structure function $F_{2}$ and its
derivative. Then we consider the positive roots for the non-linear
corrections to the gluon distribution in Eq.(16). For
$Q^{2}<2~\mathrm{GeV}^{2}$ in the range $10^{-5}<x<10^{-1}$, the
positive roots are dominant in Eq.(16).
\begin{figure}[h]
\includegraphics[width=0.45\textwidth]{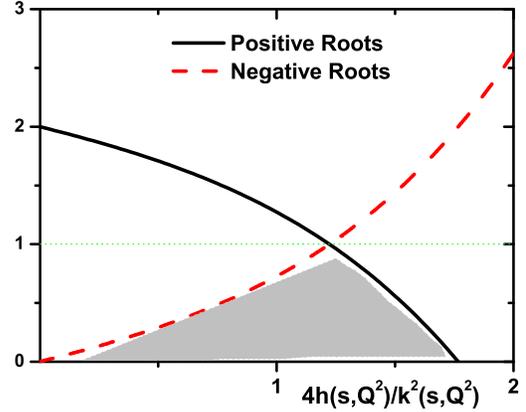}
\caption{The shaded area represents the convergence of positive
and negative roots in Eq.(16).}\label{Fig1}
\end{figure}
In this domain, the non-linear gluon distribution is defined by
\begin{eqnarray}
G^{\mathrm{NLC}}(x,Q^{2})&=&
{\mathcal{L}}^{-1}[k(s,Q^{2})-\frac{h(s,Q^{2})}{k(s,Q^{2})}-\frac{h^{2}(s,Q^{2})}{k^{3}(s,Q^{2})}\nonumber\\
&&-2\frac{h^{3}(s,Q^{2})}{k^{5}(s,Q^{2})}+...;\nu]\nonumber\\
&&=\frac{2\alpha_{s}}{\pi{\zeta}}\int_{x}^{1}\frac{dy}{y}[(\frac{x}{y})-2(\frac{x}{y})^{2}+2(\frac{x}{y})^{3}]\nonumber\\
&&-\mathrm{Eq}.(10)-{\mathcal{L}}^{-1}[\frac{h^{2}(s,Q^{2})}{k^{3}(s,Q^{2})}\nonumber\\
&&+2\frac{h^{3}(s,Q^{2})}{k^{5}(s,Q^{2})}+...;\nu].
\end{eqnarray}
In next section, we will describe the numerical results for
$G^{\mathrm{NLC}}(x,Q^{2})$ by using the analytical solutions,
(i.e., Eqs.(18) and (19), at low and high $Q^{2}$ values.\\

\subsection{IV. Results and Discussions}
Now we use the LO approximation of $\alpha_{s}(Q^{2})$ which is
defined by
$$
\alpha_{s}(Q^{2})=\frac{4\pi}{(11-\frac{2}{3}n_{f}){\ln}(Q^{2}/\Lambda^{2})},
$$
with $n_{f}=5$ and $\Lambda_{5}=80.80~\mathrm{MeV}$ for $Q>m_{b}$,
$n_{f}=4$ and $\Lambda_{4}=136.80~\mathrm{MeV}$ for
$m_{c}<Q<m_{b}$, and $n_{f}=3$ and
$\Lambda_{3}=136.80~\mathrm{MeV}$ for $Q<m_{c}$ where $\Lambda
^{,}$s have been extracted with $\alpha_{s}(m_{z}^{2})=0.1166$ at
the $Z$-boson mass. By using the Eqs.(11), (18) and (19), we can
extract numerically the linear and non-linear gluon distribution
inside the proton from the parameterization of the proton
structure function and its derivatives. Then compare them with the
results in Refs.[11] and [12].\\
The results of the calculations based on the linear gluon
distribution (i.e., Eq.(11)) are shown in Fig.3. In this figure
the straight lines represent the solutions resulted from the
Laplace transform technique. We take the $n_{f}=3$ for
$Q<1.3~\mathrm{GeV}$, $n_{f}=4$ for $Q<4.5~\mathrm{GeV}$ and
$n_{f}=5$ for $Q>4.5~\mathrm{GeV}$ as the gluon distribution
function $G(x,Q^{2})$ depends on $n_{f}$. In Fig.4, the results
for the linear gluon distributions have been shown and compared
with the parametrization methods in Refs.[11,12] for $Q^{2}=10$
and $100~\mathrm{GeV}^{2}$. For $Q^{2}=10~\mathrm{GeV}^{2}$ these
results compared with $G_{n_{f}=4}(x,Q^{2})$ in Refs.[11] and [12]
\footnote{ In Ref.[11] the gluon distribution for $n_{f}=4$ is
just $\frac{3}{5}G_{n_{f}=3}(x,Q^{2})$, where
$G_{n_{f}=3}(x,Q^{2})$ is obtained from a fit to ZEUS data [13]
into an expression in both $\ln(Q^{2})$ and $\ln(1/x)$ to include
the effects of heavy-quark masses. In Ref.[12] authors obtained an
analytical solution for $G(x,Q^{2})$ using a Froissart bounded
structure function for $0<x{\lesssim}0.09$. Those obtained a
simple quadratic polynomial in $\ln(1/x)$ with quadratic
polynomial coefficients in $\ln(Q^{2})$.}. For
$Q^{2}=100~\mathrm{GeV}^{2}$ where the active flavor is $n_{f}=5$,
we compared our results with those obtained in
Ref.[11]\footnote{In Ref.[11] authors obtained the gluon
distribution $G(x,Q^{2})$ for 5 active quarks ( for massless $u$,
$d$, $s$ and massive $c$, $b$ quarks) into the massless gluon
distribution $G_{n_{f}=3}(x,Q^{2})$, as
$G_{n_{f}=5}(x,Q^{2})=\frac6{11}G_{n_{f}=3}(x,Q^{2})$. Also
authors obtained an excellent fit to the gluon distribution for
$n_{f}=5$ using a quadratic expression in $\ln{1/x}$ and a much
more complicated power series in $\ln(Q^{2})$ for
$x{\lesssim}0.05$.}. This figure indicate that the obtained
results from the present analysis, based on the Laplace transform
technique using the number of active flavors, are in good
agreements with the
ones obtained from the parametrization methods.\\
In Fig.5 we plot the $Q^{2}$ dependence of the non-linear
corrections to the gluon distribution for
$\mathcal{R}=2~\mathrm{GeV}^{-1}$ at some representative $x$ and
check the compatibility of the non-linear results with the linear
gluon distributions. In this figure (i.e., Fig.5) we can see that
at large scales, the nonlinear corrections (due to the $1/Q^{2}$
dependence) relax into the linear gluon distribution functions.
Also the non-linear corrections play an important role on gluon
distribution as $x$ and $Q^{2}$ decreases. In Fig.5 we observe
that the dot line separates linear and non-linear behaviors of the
gluon distributions. It is seen that these distributions change
discontinuously at each threshold but remain constant between
thresholds. These results are comparable with
Eskola-Honkanen-Kolhinen-Qiu-Salgado (EHKQS), which obtained
[14-15] the parton distribution functions using the CTEQ6L [16]
with respect to the non-linear GLRMQ evolution equations.\\
At low scales $Q^{2}<2~\mathrm{GeV}^{2}$, where the positive roots
are dominate in the non-linear distributions, we obtained the
non-linear behavior of the gluon distribution in Fig.6. In this
figure (i.e., Fig.6), we show the non-linear corrections to the
gluon distribution determined from Eqs.(18) and (19) as a function
of $x$ for two different values of $Q$, namely
$Q=1.14~\mathrm{GeV}$ and $1.30~\mathrm{GeV}$ with respect to the
Lpalce transforms method. In these calculations, we use the
non-linear gluon distribution with $n_{f}=3$ for
$Q=1.14~\mathrm{GeV}$ and compared the obtained results with the
parametrization method in Ref.[11] at the same active flavor
number. Also the charm threshold is shown for
$Q=1.30~\mathrm{GeV}$ with $n_{f}=3$ and $4$ in Fig.6 and compared
the obtained results with the parametrization method in Ref.[11]
at each active flavor number. A depletion occurs at
$x{\leq}10^{-4}$ where these results show that the non-linear
behavior of the gluon distribution function is tamed with respect
to the positive roots. This taming behavior of non-linear gluon
distribution function towards low $x$ at low $Q^{2}$ values become
significant at the hot spot point. Further the computed values of
the gluon distribution with non-linear effects play an
increasingly important role at low
$x$ and low $Q^{2}$ values.\\
In conclusion, we have presented a certain theoretical model to
describe the non-linear corrections to the gluon distribution
function based on the Laplace transforms method  at low values of
$x$ and $Q^{2}$ in a limited approach. A detailed analysis has
been performed to find an analytical solution of the linear and
non-linear gluon distribution functions from the proton structure
function and its derivative. The effect of non-linear corrections
on the behavior of $G(x,Q^{2})$ with decreasing $Q^{2}$ become
significant at the hot spot point. At high $Q^{2}$ values the
non-linear corrections relax into the linear gluon distribution
function. The nonlinear corrections have been tamed the behavior
of the gluon distribution function
at $Q^{2}<2~\mathrm{GeV}^{2}$ in comparison with the linear behavior.\\

\subsection{ACKNOWLEDGMENTS}

I thank the respectable referee for giving the main idea of this
work. I sincerely thank the referee for his/her invaluable
comments during the review process one of them my papers in EPJC.
 The feedback from the referee was very important for me.\\

\subsection{Appendix A}

The explicit expression for the proton structure function
suggested in Ref.[10] is defined by the following form
\begin{eqnarray}
F^{\gamma p}_{ 2}(x,Q^{2})& =& D(Q^{2})(1-
x)^{n}[C(Q^{2})+A(Q^{2})\ln(\frac{1}{x}\frac{Q^{2}}{Q^{2}+\mu^{2}})\nonumber\\
&&+B(Q^{2})\ln^{2}(\frac{1}{x}\frac{Q^{2}}{Q^{2}+\mu^{2}})],
\end{eqnarray}
where
\begin{eqnarray}
 A(Q^{2})& =& a_{0} + a_{1} {\ln}(1+\frac{Q^{2}}{\mu^{2}}) + a_{2}{\ln}^{2}(1+\frac{Q^{2}}{\mu^{2}})
 ,\nonumber\\
B(Q^{2})& =& b_{0} + b_{1} {\ln}(1+\frac{Q^{2}}{\mu^{2}}) +
b_{2}{\ln}^{2}(1+\frac{Q^{2}}{\mu^{2}})
 ,\nonumber\\
C(Q^{2})& =& c_{0} + c_{1}
{\ln}(1+\frac{Q^{2}}{\mu^{2}}),\nonumber\\
D(Q^{2})& =& \frac{Q^{2}(Q^{2}+\lambda M^{2})}{(Q^{2}+M^{2})^2}.
\end{eqnarray}
Here $M$ is the effective mass and $\mu^{2}$ is a scale factor.
The additional parameters with their statistical errors are given
in Table I.\\
\begin{table}[h]
\caption{ The effective parameters at low $x$ for
$0.15~\mathrm{GeV}^{2}<Q^{2}<3000~\mathrm{GeV}^{2}$ provided by
the following values. The fixed  parameters are defined by the
Block-Halzen fit to the real photon-proton cross section as
$M^{2}=0.753 \pm 0.068~ \mathrm{GeV}^{2}$, $\mu^2 = 2.82 \pm
0.290~ \mathrm{GeV}^{2}$ and $c_{0} = 0.255 \pm 0.016$ [10].}
\begin{tabular} {cccc}
\toprule \\  \multicolumn{2}{c}{parameters \quad \quad \quad ~~~~~~~~~~~~~~~~value}    \\ &&&\\ \hline \\ &&&\\
  $a_{0} $  &   \quad  $8.205\times 10^{-4}~~  \pm  4.62\times10^{-4} $  \\

  $a_{1} $  &   \quad   $-5.148\times 10^{-2}\pm 8.19\times10^{-3}$  \\

  $a_{2}$   &    \quad  $-4.725\times 10^{-3}\pm 1.01\times10^{-3}$   \\  &&&\\

 $b_{0}$   &   \quad   $2.217\times 10^{-3}\pm 1.42\times10^{-4} $ \\

 $b_{1}$   &   \quad   $1.244\times 10^{-2}\pm 8.56\times10^{-4}$  \\

 $b_{2}$    &    \quad  $5.958\times 10^{-4}\pm 2.32\times10^{-4} $ \\ &&& \\

$c_{1}$& \quad  $1.475\times 10^{-1}~\pm 3.025\times10^{-2}$ & &\\

$n$& \quad  $11.49\pm 0.99$ & &\\

$\lambda$& \quad  $2.430~\pm 0.153$ & &\\

$\chi^{2}(\mathrm{goodness~ of~ fit})$ &  \quad  $0.95$ & &\\
\hline

\end{tabular}
\end{table}
\begin{figure}
\includegraphics[width=0.55\textwidth]{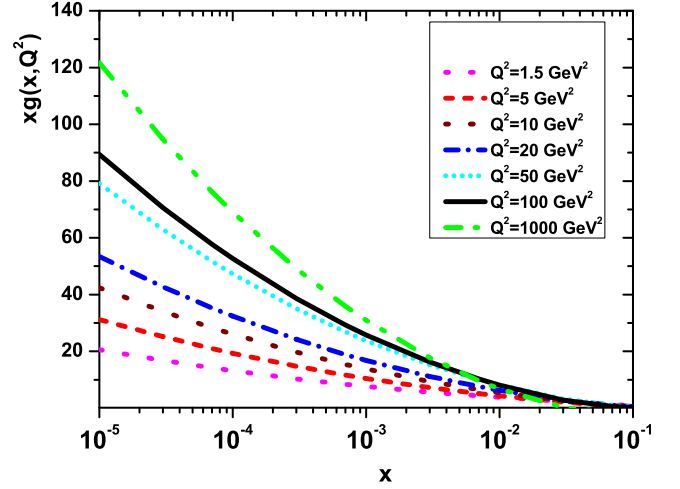}
\caption{Our results for the linear gluon distribution, using
Eq.(10), as a function of $x$ in a wide range of $Q^{2}$ values,
$1~\mathrm{GeV}^{2}<Q^{2}{\leq}1000~\mathrm{GeV}^{2}$.
}\label{Fig3}
\end{figure}
\begin{figure}
\includegraphics[width=0.55\textwidth]{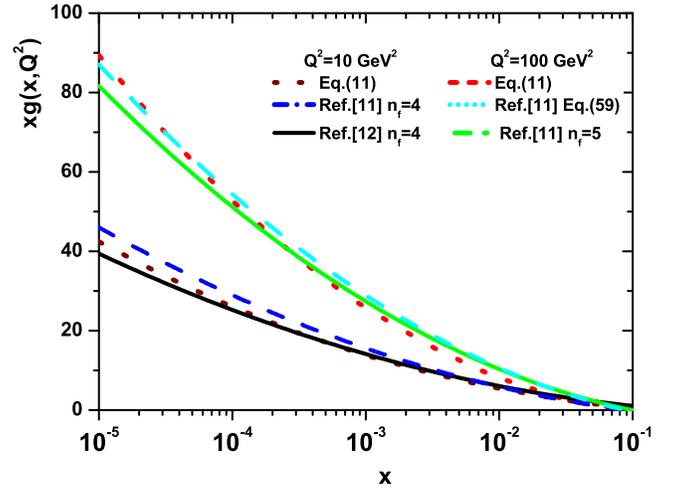}
\caption{Comparison of the $G(x,Q^{2})$ obtained from Eq.(11),
with the parametrization methods in Refs. [11,12] at $Q^{2}=10$
and $100~\mathrm{GeV}^{2}$.}\label{Fig4}
\end{figure}
\begin{figure}
\includegraphics[width=0.55\textwidth]{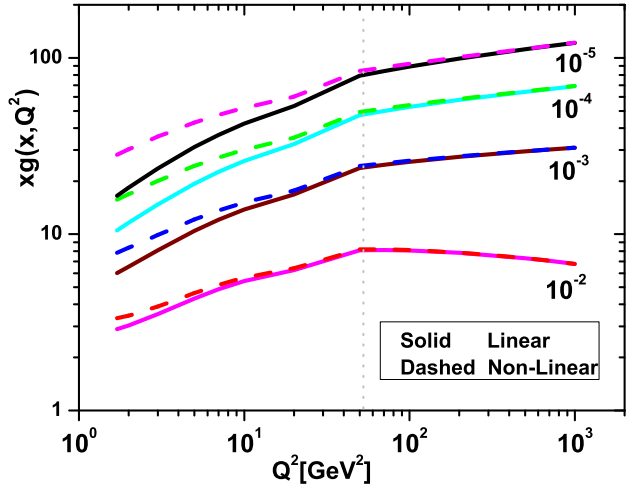}
\caption{Linear and non-linear behavior of the gluon distributions
for several fixed values of $x$ in a wide range of $Q^{2}$ shows
that the effect of the nonlinear terms vanishes as $Q^{2}$
increases.}\label{Fig5}
\end{figure}
\begin{figure}
\includegraphics[width=0.55\textwidth]{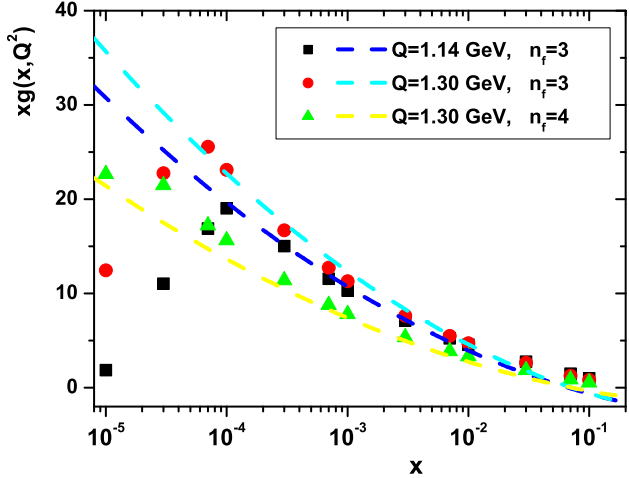}
\caption{Non-linear gluon distribution at low $Q$ values due to
the positive and negative roots in Eqs.(18) and (19) compared with
the linear gluon distribution functions (dashed lines) obtained in
Ref.[11] at each active flavor number. }\label{Fig6}
\end{figure}
\section{References}
1. L.N. Lipatov, Sov. J. Nucl. Phys.{\bf20}, 94 (1975); V.N.
Gribov, L.N. Lipatov, Sov. J. Nucl. Phys.{\bf15}, 438 (1972); G.
Altarelli, G. Parisi, Nucl. Phys. B{\bf126}, 298 (1977); Yu.L.
Dokshitzer, Sov. Phys.
JETP {\bf46}, 641 (1977).\\
2. H.Khanpour, A.Mirjalili and S.Atashbar Tehrani,
Phys.Rev.C{\bf95}, 035201 (2017); H.Khanpour, M.Goharipour and
V.Guzey, Eur.Phys.J.C{\bf78}, 7(2018); S. Mohammad Moosavi Nejad,
H.Khanpour, S.Atashbar Tehrani and M.Mahdavi, Phys.Rev.C{\bf94},
045201 (2016).\\
3. G.R.Boroun, S.Zarrin and F.Teimoury, Eur.Phys.J.Plus {\bf130},
214(2015); F.Teimoury Azadbakht and G.R.Boroun,
Int.J.Theor.Phys.{\bf57}, 495 (2018); S.Zarrin and G.R.Boroun,
Nucl.Phys.B{\bf922}, 126(2017); F.Teimoury Azadbakht, G.R.Boroun
and B.Rezaei,
Int.J.Mod.Phys.E{\bf27}, 1850071 (2018).\\
4. M.Mottaghizadeh, F.Taghavi Shahri and P.Eslami,
Phys.Lett.B{\bf773}, 375(2017); M.Mottaghizadeh, P.Eslami and
F.Taghavi-Shahri, Int.J.Mod.Phys.A{\bf32}, 1750065(2017); S.Dadfar
and S.Zarrin,
Eur.Phys.J.C{\bf80}, 319(2020); H.Hosseinkhani, M.Modarres and N.Olanj,
 Int.J.Mod.Phys.A{\bf32}, 1750121 (2017).\\
5. M. M. Block, L. Durand, and D. W. McKay, Phys. Rev. D {\bf79},
014031 (2009).\\
6. LHeC Collaboration and FCC-he Study Group , P.Agostini et al.,
CERN-ACC-Note-2020-0002, arXiv:2007.14491 [hep-ex] (2020).\\
7. L.V.Gribov, E.M.Levin and M.G.Ryskin, Phys.Rept.{\bf100}, 1
(1983); A.H.Mueller and J.w.Qiu, Nucl.Phys.B {\bf268},
427 (1986).\\
8. A. V. Kotikov and V. N. Velizhanin, arXiv: 0501274 [hep-ph]
(2005).\\
9. F. D. Aaron et al. (H1 and ZEUS Collaborations), JHEP
{\bf1001}, 109 (2010).\\
10. M. M. Block, L. Durand and P. Ha, Phys. Rev. D {\bf89}, 094027
(2014).\\
11. Martin M.Block and L.Durand, arXiv:0902.0372 [hep-ph] 2009.\\
12. M. M. Block, L. Durand, and D. W. McKay, Phys. Rev. D {\bf77},
094003 (2008).\\
13 J. Breitweg et al. (ZEUS), Phys. Lett. B {\bf487}, 53 (2000);
S. Chekanov et al. (ZEUS), Eur. Phys. J. C {\bf21}, 443 (2001).\\
14. K. J. Eskola, H. Honkanen, V. J. Kolhinen, J.-W. Qiu and C. A.
Salgado, Nucl. Phys. B {\bf660}, 211 (2003).\\
15. A. Dainese et al., HERA-LHC Workshop, DESY,  (2005).\\
16. J. Pumplin et al., JHEP {\bf07}, 012 (2002).\\

\end{document}